\documentclass[conference]{IEEEtran}
\IEEEoverridecommandlockouts

\usepackage{cite}
\usepackage{amsmath,amssymb,amsfonts}
\usepackage{algorithmic} 
\usepackage{graphicx} 
\usepackage{textcomp} 
\usepackage{xcolor}  
\usepackage{booktabs} 
\usepackage{multirow}
\usepackage{url}
\usepackage{hyperref}
\hypersetup{colorlinks=true, linkcolor=black, citecolor=black, urlcolor=blue}

\def\BibTeX{{\rm B\kern-.05em{\sc i\kern-.025em b}\kern-.08em
    T\kern-.1667em\lower.7ex\hbox{E}\kern-.125emX}} 

\begin{document}

\title{Concurrent Streaming, Viewer Transfers, and Audience Loyalty in a Creator Ecosystem: A Minute-Level Longitudinal Study}

\author{
\IEEEauthorblockN{Maxwell Shepherd}
\IEEEauthorblockA{Johns Hopkins University\\
\texttt{msheph15@jh.edu}}
}

\maketitle

\begin{abstract}
Live streaming platforms host interconnected communities of content creators whose audiences overlap and interact in ways that are poorly understood at fine temporal resolution. We present a descriptive longitudinal study of audience behavior within a creator ecosystem, analyzing 2.9 million minute-by-minute viewership observations across 7,762 livestreams from 18 affiliated channels over 3.3 years. We find that (1) concurrent streaming is associated with substantial raw per-stream audience decreases (14,377 to 6,057 viewers as concurrent stream count rises from 1 to 9), though hour-of-day controls reduce the residualized correlation to $\rho = -0.165$, indicating that scheduling confounds account for much of the observed drop; (2) algorithmically detected viewer transfer events achieve a median efficiency of approximately 50\% across 3,243 candidate events; and (3) audience loyalty metrics (stability, competition resistance, retention, and floor ratio) vary substantially across creators within the same organization, with competition resistance ranging from 0.36 to 1.00, indicating that audience exclusivity is a creator-level rather than organization-level property. These findings provide practical benchmarks for creator organizations making scheduling, cross-promotion, and talent management decisions.
\end{abstract}

\begin{IEEEkeywords}
livestreaming, audience overlap, creator ecosystems, concurrent streaming, viewer transfer, audience loyalty, Hololive, YouTube, virtual YouTuber
\end{IEEEkeywords}

\section{Introduction}

Live streaming has become a major mode of online media production and audience participation, with Twitch alone recording billions of hours watched in 2024 and creator-centered live platforms sustaining large viewer communities at global scale~\cite{b1,b2}. A distinctive feature of this landscape is the emergence of creator networks in which affiliated streamers collaborate, cross-promote, and sometimes compete for viewer attention within a common platform~\cite{b2}.

A practical question for both platform operators and creator organizations is whether concurrent streaming among affiliated creators hurts individual streamers' audiences. The intuition that concurrent streams ``split the audience'' is widely held and influences scheduling decisions, yet it has rarely been tested with fine-grained longitudinal data. Similarly, the effectiveness of viewer transfer mechanisms (colloquially called ``raids'') and the degree to which audiences are exclusive to individual creators are poorly quantified.

Despite growing academic interest in live streaming~\cite{b4,b5}, much prior work has focused on ethnography, surveys, short-window observational datasets, or chat-centered participation measures rather than multi-year minute-by-minute audience panels~\cite{b4,b5,b6}.

In this paper, we use a dataset of 2,935,985 minute-by-minute viewership records spanning 7,762 livestreams from 18 channels over 3.3 years to address three questions:

\begin{enumerate}
    \item \textbf{Does concurrent streaming hurt individual creators?} We measure per-stream audience dilution when multiple affiliated creators stream simultaneously, controlling for time-of-day confounds.
    \item \textbf{How effective are viewer transfers?} We algorithmically detect viewer transfer events and quantify the fraction of a stream's audience that transfers to a concurrent ecosystem stream.
    \item \textbf{How much do audiences vary in exclusivity?} We decompose audience loyalty into four components (stability, competition resistance, post-peak retention, and floor ratio), revealing dramatic variation across creators within the same organization.
\end{enumerate}

\section{Related Work}

\subsection{Live Streaming Research}

The live streaming domain has attracted increasing research attention. Hamilton et al.~\cite{b4} examined community formation on Twitch, finding that streamers cultivate distinct community identities. Deng et al.~\cite{b8} analyzed 11 months of Twitch activity and showed that the platform is highly skewed, with a small number of games dominating attention and live events contributing to large traffic spikes. Haimson and Tang~\cite{b3} examined what makes remote live events engaging across Facebook Live, Periscope, and Snapchat, identifying four dimensions of engagement: immersion, immediacy, interaction, and sociality.

Recent work has begun to address inter-stream dynamics. Jia et al.~\cite{b10} characterized the creators and spectators of game replays and live streaming, providing foundational understanding of streamer--viewer relationships. Wu et al.~\cite{b11} studied cooperation between content creators on Twitch, finding that different cooperative behaviors are associated differently with creator performance rather than uniformly benefiting growth. However, these studies primarily examine creator--viewer relations, platform structure, or cooperation behaviors rather than minute-level longitudinal audience dynamics within a single affiliated creator ecosystem.

\subsection{Network Analysis of Creator Ecosystems}

Our analysis also connects to broader work on competition and differentiation in media markets, which shows that audience demand can shape how media producers position themselves relative to rivals~\cite{b12}. Borghol et al.~\cite{b13} studied content-agnostic factors affecting YouTube video popularity, finding that extrinsic factors beyond content quality substantially influence viewership outcomes. Ribeiro et al.~\cite{b14} tracked user intersection and migration across YouTube ideological communities, illustrating how behavioral traces can be used to study cross-community audience movement.

\subsection{Virtual YouTubers (VTubers)}

Research on VTubers remains limited, so we draw on both VTuber-specific work and adjacent live-streaming studies. Lu et al.~\cite{b17} examined live streaming practices in China, highlighting the roles of gifting, social interaction, and fan-group dynamics, while Lu et al.~\cite{b18} specifically studied how viewers engage with and perceive virtual YouTubers. Our dataset focuses on Hololive Production, one of the largest VTuber agencies, whose organizational structure creates a natural laboratory for studying creator ecosystem dynamics.

\section{Dataset and Methodology}

\subsection{Dataset Description}

Our dataset comprises minute-by-minute viewer counts collected from YouTube Live streams of Hololive English (HoloEN) creators. Table~\ref{tab:dataset} summarizes the dataset characteristics.

\begin{table}[htbp]
\caption{Dataset Summary}
\begin{center}
\begin{tabular}{lr}
\toprule 
\textbf{Metric} & \textbf{Value} \\
\midrule
Total observations & 2,935,985 \\
Unique streams & 7,762 \\
Unique channels & 18 \\
Observation period & Nov 2022 -- Mar 2026 \\
Temporal resolution & 1 minute \\
Median stream duration & 181.2 minutes \\
Mean viewers per stream & 7,587 \\
Max observed viewers (single stream) & 243,897 \\
\bottomrule
\end{tabular}
\label{tab:dataset}
\end{center}
\end{table}

The 18 channels span four organizational ``generations'' (debut cohorts), enabling analysis of how cohort structure affects audience overlap. Stream metadata includes channel identity, video identifier, scheduled and actual start times, stream title, and minute-level viewer counts.

\subsection{Channel Activity Disparities}

The channels in our dataset exhibit substantial disparities in streaming activity. Streaming frequency ranges from 3.1 streams per month (Gura, 90 streams) to 22.8 streams per month (Bijou, 720 streams), and active periods range from 632 days (Justice generation, debuting most recently) to up to 1,220 days for the earliest channels.

\subsection{Audience Overlap Estimation}

Since we observe aggregate viewership counts rather than individual viewer identities, we estimate audience overlap indirectly through \textit{viewership transfer analysis}. When stream $A$ begins while stream $B$ is ongoing, a drop in $B$'s viewership attributable to $A$'s start provides evidence of shared audience. We use only stream \textit{start} events (not end events) to avoid confounding with natural viewership decay. All stream start events are processed exhaustively (no sampling), and the concurrent stream $B$ must span the full measurement window (having started at least $\delta$ minutes before and ending at least $\delta$ minutes after the event) to ensure stable baseline measurement.

Formally, for a stream start event at time $t_0$, we define the viewership change in concurrent stream $B$ as:

\begin{equation}
\Delta V_B = \bar{V}_B^{[t_0, t_0+\delta]} - \bar{V}_B^{[t_0-\delta, t_0]}
\label{eq:transfer}
\end{equation}

\noindent where $\bar{V}_B^{[a,b]}$ denotes the mean viewership of stream $B$ over interval $[a,b]$ and $\delta = 8$ minutes. The estimated overlap between channels $i$ and $j$ is the median normalized transfer across all observed start events:

\begin{equation}
O_{ij} = \text{median}\left\{\frac{-\Delta V_j}{\bar{V}_j^{[t_0-\delta, t_0]}} \bigg| \text{stream}_i \text{ starts during stream}_j\right\}
\label{eq:overlap}
\end{equation}

\noindent The pairwise overlap matrix is then symmetrized: $\hat{O}_{ij} = (O_{ij} + O_{ji})/2$. We validate this methodology via permutation testing (Section~\ref{sec:validation}).

\subsection{Viewer Transfer Detection}

A \textit{viewer transfer event} occurs when a substantial portion of one stream's audience moves to another active stream near the first stream's conclusion. This may result from an intentional directive by the ending streamer (colloquially called a ``raid'' or ``host'') or from viewers independently seeking another stream within the ecosystem. We detect these events algorithmically by exhaustively scanning all stream end events: for each stream $A$ ending at time $t_e$, we examine every concurrent stream $B$ (which must have started at least 5 minutes before and end at least 5 minutes after $t_e$). We compare the mean viewership of $B$ in the 3 minutes before $t_e$ to the peak viewership of $B$ in the 5 minutes after $t_e$. A candidate transfer event is flagged when:

\begin{enumerate}
    \item The spike exceeds 10\% of $B$'s pre-event average viewership.
    \item The spike exceeds 100 absolute viewers.
    \item The spike exceeds 5\% of $A$'s stream-average viewers (to exclude coincidental fluctuations).
    \item Stream $A$ ended with at least 200 viewers (to exclude abandoned streams).
\end{enumerate}

\noindent Transfer efficiency is defined as the spike magnitude divided by stream $A$'s final viewer count. We estimate approximately 5\% of detected events are false positives (coincidental spikes unrelated to the ending stream), based on the fraction of events with efficiency exceeding 100\%.

\subsection{Loyalty Component Construction}

We measure four dimensions of audience exclusivity for each channel:

\begin{itemize}
    \item \textbf{Stability ($S$)}: $1 - \text{CV}/2$ where CV is the coefficient of variation of stream-average viewership, clipped to $[0,2]$, capturing how consistent a channel's audience size is across streams.
    \item \textbf{Competition Resistance ($R$)}: Ratio of mean viewership during streams with concurrent ecosystem competition to mean viewership during solo streams, clipped to $[0,1]$. Higher values indicate audiences that are retained regardless of concurrent alternatives.
    \item \textbf{Post-Peak Retention ($P$)}: Median fraction of peak viewership remaining at the midpoint of the post-peak segment of a stream. This measures how gradually audiences depart after peak engagement.
    \item \textbf{Floor Ratio ($F$)}: 10th-percentile stream-average viewership divided by mean stream-average viewership, capturing how much a channel's audience shrinks during its least popular streams relative to its typical performance.
\end{itemize}

Each component is independently informative and constitutes the primary unit of analysis. For visual reference, we also report a weighted composite $L = 0.30S + 0.25R + 0.25P + 0.20F$, but we caution that the weights are subjective and different weightings would produce different channel rankings. The key findings concern variation in individual components, not composite scores.

\subsection{Permutation Validation}
\label{sec:validation}

To test whether the observed frequency of concurrent-at-start events between channel pairs exceeds what would occur under independent random scheduling, we conduct a permutation test with 1,000 iterations. In each iteration, we randomly reassign stream start times uniformly within the observation window for each channel (preserving the number of streams and their durations per channel) and count the number of concurrent-at-start events per channel pair under the shuffled schedule. We compare the observed pairwise event counts to their null distributions, finding that 53\% of channel pairs (81 of 153 tested) exhibit concurrent-at-start event frequency significantly exceeding the uniform null at $p < 0.01$. We note that this null model does not preserve hour-of-day or day-of-week structure, so some of the excess concurrency may reflect shared scheduling patterns (e.g., peak-hour clustering) rather than deliberate coordination. The test validates that concurrent streaming is non-random, but does not by itself confirm audience-level overlap.

\section{Results}

\subsection{Audience Overlap Structure}

\begin{figure}[htbp]
\centerline{\includegraphics[width=0.98\columnwidth]{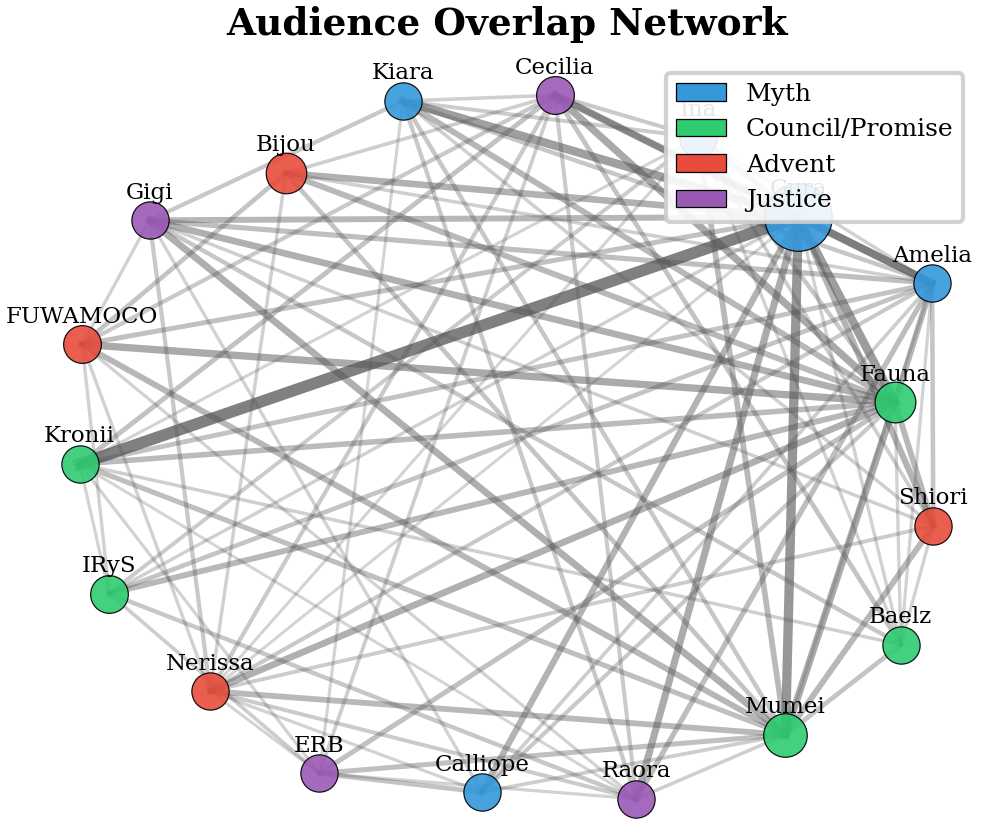}}
\caption{Audience overlap network. Node size proportional to average viewership; edge weight proportional to estimated overlap. Colors indicate organizational generation.}
\label{fig:network}
\end{figure}

Fig.~\ref{fig:network} displays the audience overlap network, where nodes represent channels (sized by average viewership) and edges represent estimated audience overlap (weighted by overlap strength). The network exhibits structure partially corresponding to organizational generations, with cross-generation bridges connecting otherwise separate audience clusters.

\begin{figure}[htbp]
\centerline{\includegraphics[width=0.95\columnwidth]{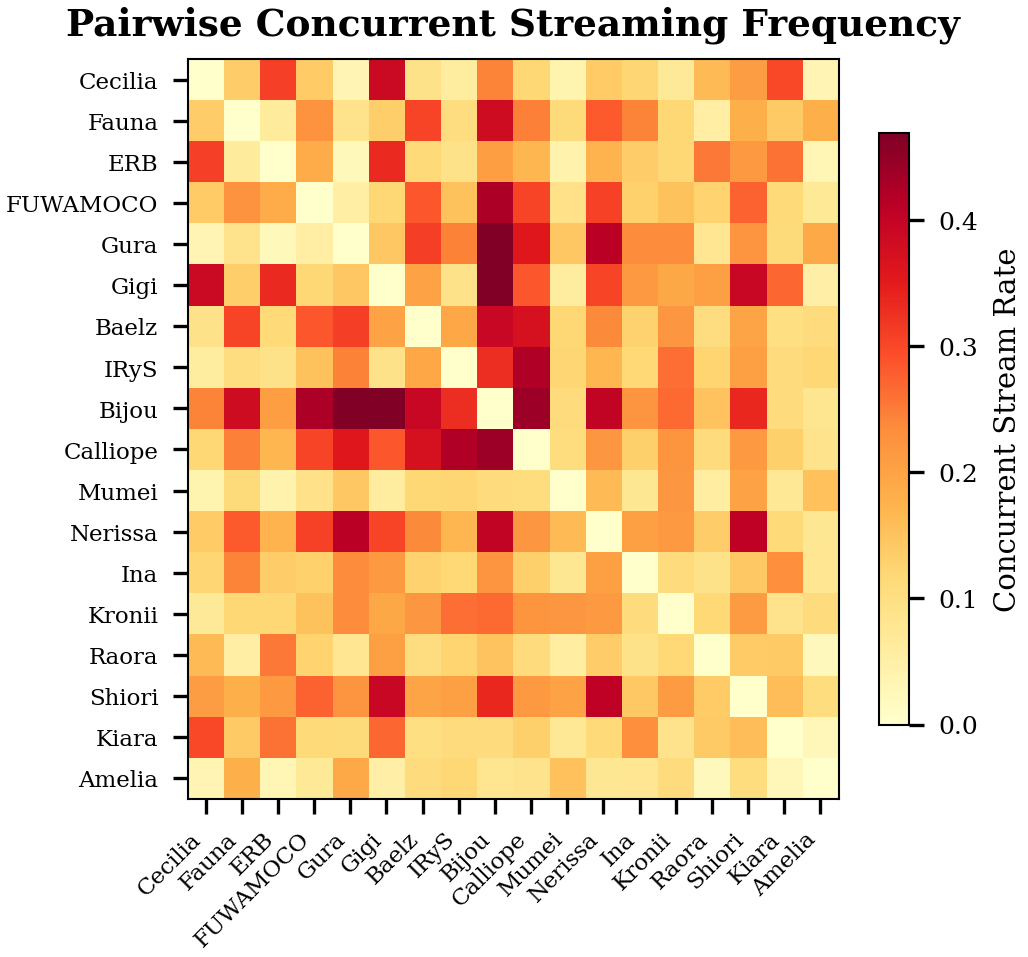}}
\caption{Pairwise concurrent streaming frequency. Darker cells indicate higher rates of temporal overlap between channels.}
\label{fig:heatmap}
\end{figure}

Fig.~\ref{fig:heatmap} shows the full pairwise concurrent streaming frequency matrix, revealing that overlap is highly non-uniform: certain channel pairs stream concurrently far more often than others, likely driven by scheduling proximity rather than organizational grouping alone.

Among a constant set of six channels active for nearly the entire study period (Baelz, Calliope, IRyS, Ina, Kiara, Kronii), pairwise concurrent streaming frequency shows a statistically significant upward trend ($\rho = 0.63$, $p < 0.001$), though the absolute change is small (from 8.4\% to 8.9\% over 3.3 years).

\subsection{Per-Stream Dilution Under Concurrent Streaming}

\begin{figure*}[htbp]
\centerline{\includegraphics[width=0.95\textwidth]{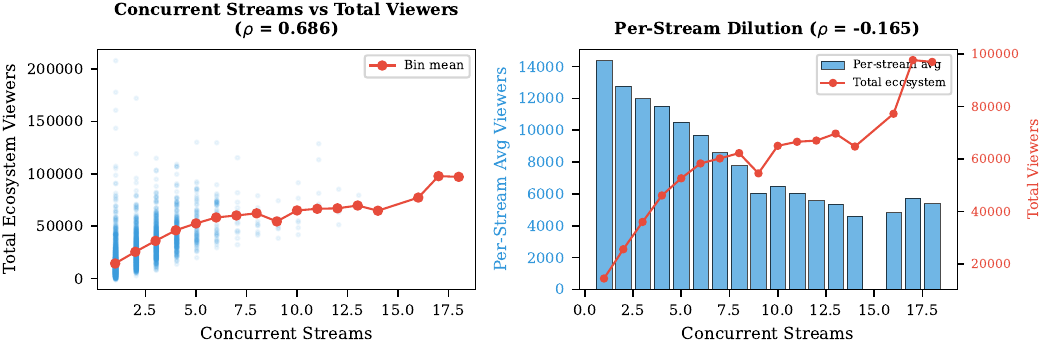}}
\caption{Concurrent streaming and per-stream dilution (HoloEN channels only, minute-level). Left: total ecosystem viewership vs. concurrent stream count ($\rho = 0.686$), largely reflecting scheduling endogeneity. Right: per-stream average viewership (bars) and total ecosystem viewership (line) by concurrent stream count. Per-stream averages drop substantially with concurrent stream count, though the residualized within-hour effect is smaller ($\rho = -0.165$).}
\label{fig:dilution}
\end{figure*}

Fig.~\ref{fig:dilution} presents the relationship between concurrent stream count and viewership among HoloEN channels, computed at minute-level resolution across 903,778 observations. The left panel shows a positive correlation between concurrent stream count and total ecosystem viewership ($\rho = 0.686$); however, this correlation is largely an artifact of scheduling endogeneity, as streamers are more likely to stream during hours when viewers are available, so both concurrent streams and total viewership are high during peak hours.

The more informative result is in the right panel: per-stream average viewership drops substantially as concurrent stream count rises, from 14,377 viewers at 1 concurrent stream to 6,057 at 9 concurrent streams. This raw decrease reflects two overlapping factors: audience splitting across concurrent streams, and scheduling confounds (more streams are live during peak hours, so higher-concurrency buckets sample from different time periods than lower-concurrency buckets). After controlling for hour-of-day effects via residualization, the Spearman rank correlation between hour-residualized concurrent stream count and hour-residualized per-stream viewership is $\rho = -0.165$ (day-level block bootstrap 95\% CI: $[-0.186, -0.145]$, $p < 0.001$), indicating that within-hour competition accounts for a small but statistically significant portion of the overall drop. Because consecutive minute-level observations are temporally autocorrelated, we use a block bootstrap (resampling 1,209 calendar days with replacement, 2,000 iterations) rather than the naive iid $p$-value to ensure honest inference.

From a scheduling perspective, the raw per-stream decrease is substantial regardless of its cause: a creator streaming alone averages 14,377 viewers, while a creator streaming alongside 8 others averages 6,057. The hour-of-day residualization isolates the within-hour competition effect as small ($\rho = -0.165$), but does not diminish the practical reality that streaming during high-concurrency periods is associated with significantly lower per-stream viewership.

\subsection{Viewer Transfer Events}

\begin{figure}[htbp]
\centerline{\includegraphics[width=0.95\columnwidth]{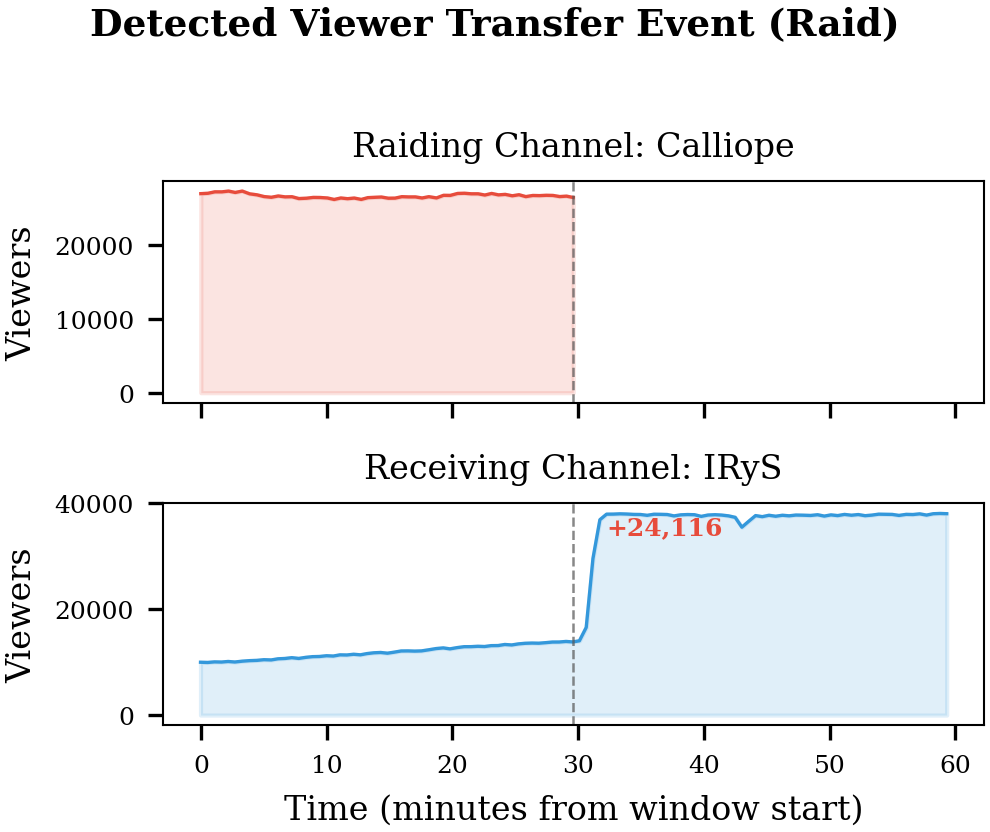}}
\caption{Detected viewer transfer event: Calliope $\rightarrow$ IRyS on 2025-06-21. Top panel: ending channel viewership. Bottom panel: receiving channel with +24,116 viewer spike at transfer time (dashed line).}
\label{fig:transfer}
\end{figure}

Our exhaustive transfer detection algorithm identifies 3,243 candidate viewer transfer events across all stream endings in the dataset (Fig.~\ref{fig:transfer}). We estimate approximately 5\% of these are false positives (coincidental viewership spikes unrelated to the ending stream), based on the fraction of events with efficiency exceeding 100\%.

Among events with plausible efficiency ($\leq$100\%), the mean spike magnitude is 2,552 viewers and the median transfer efficiency is approximately 49\% (IQR: 29\%--60\%). This suggests that roughly half of a stream's departing audience transfers to a concurrent ecosystem stream, while the remainder either leaves the platform or navigates elsewhere.

The transfer network is not uniform: the top 10 channel pairs account for 16\% of all detected events, with Bijou$\rightarrow$Calliope (70 events) being the most frequent pair. This concentration reflects scheduling adjacency and established collaborative relationships, suggesting that viewer transfer frequency depends on audience familiarity with the receiving channel, which is consistent with the audience loyalty variation documented in Section~\ref{sec:loyalty}.

\subsection{Audience Loyalty Components}
\label{sec:loyalty}

\begin{figure}[htbp]
\centerline{\includegraphics[width=0.95\columnwidth]{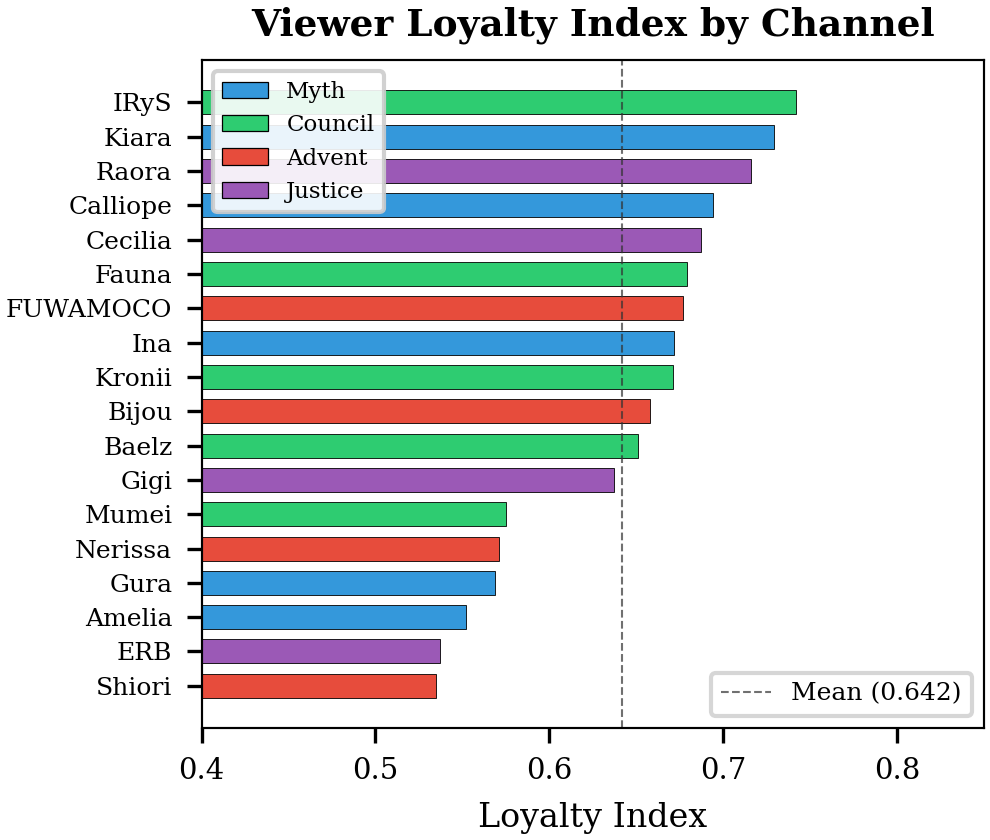}}
\caption{Audience loyalty components by channel. The composite index $L$ (bar length) is a weighted summary provided for visual reference; the individual components are the primary unit of analysis. Colors indicate organizational generation.}
\label{fig:loyalty}
\end{figure}

The four loyalty components reveal substantial heterogeneity across channels within the same organization. Table~\ref{tab:loyalty} presents the individual component values; Fig.~\ref{fig:loyalty} provides a visual summary using the composite index for ordering.

The most striking variation is in \textbf{Competition Resistance} ($R$), which ranges from 0.364 (Gura) to 1.000 (Kiara). High-$R$ channels retain nearly all their viewers when competitors stream simultaneously, while low-$R$ channels lose a substantial fraction.

\begin{table}[htbp]
\caption{Loyalty Components (Selected Channels)}
\begin{center}
\begin{tabular}{lrrrr|r}
\toprule
\textbf{Channel} & \textbf{S} & \textbf{R} & \textbf{P} & \textbf{F} & \textbf{L} \\
\midrule
IRyS     & 0.663 & 0.925 & 0.870 & 0.469 & 0.742 \\
Kiara    & 0.564 & 1.000 & 0.869 & 0.464 & 0.729 \\
Raora    & 0.483 & 0.948 & 0.930 & 0.506 & 0.716 \\
Calliope & 0.608 & 0.830 & 0.896 & 0.401 & 0.694 \\
Fauna    & 0.629 & 0.601 & 0.902 & 0.574 & 0.679 \\
Kronii   & 0.660 & 0.629 & 0.891 & 0.465 & 0.671 \\
Bijou    & 0.656 & 0.529 & 0.875 & 0.550 & 0.658 \\
Mumei    & 0.498 & 0.503 & 0.875 & 0.403 & 0.575 \\
\bottomrule
\end{tabular}
\label{tab:loyalty}
\end{center}
\vspace{-2mm}
{\footnotesize S = Stability, R = Competition Resistance, P = Post-Peak Retention, F = Floor Ratio, L = Composite Index (subjective weights: $w = (0.30, 0.25, 0.25, 0.20)$; provided for reference only).}
\end{table}

\textbf{Stability} ($S$) shows moderate variation (0.286--0.663), with some channels (ERB, Shiori) exhibiting substantially more variable audiences. \textbf{Post-Peak Retention} ($P$) is uniformly high (0.869--0.940), indicating that viewers generally stay through most of a stream once they arrive. \textbf{Floor Ratio} ($F$) varies moderately (0.373--0.574), reflecting differences in how much viewership fluctuates between a channel's most and least popular streams.

The key takeaway is that Competition Resistance, defined as the degree to which a creator's audience is retained under concurrent competition, varies by nearly 3$\times$ across creators in the same organization (0.36--1.00), while other loyalty dimensions are more uniform. This suggests that audience exclusivity is fundamentally a creator-level property, not an organization-level one.

\section{Discussion}

\subsection{Scheduling Implications}

The raw per-stream viewership drop during concurrent streaming is substantial (14,377 to 6,057), and this is the reality creators face when scheduling against peers. The residualized within-hour effect ($\rho = -0.165$) indicates that direct audience competition explains only a small portion of this drop, with most of the variation attributable to when streams are scheduled. However, the endogeneity of streaming schedules limits causal interpretation: we cannot cleanly separate audience splitting from time-of-day confounds, nor distinguish between viewers who are genuinely exclusive to one creator and viewers who would switch but are simply not present during concurrent hours.

\subsection{Viewer Transfers as Cross-Promotion}

The median 49\% transfer efficiency provides a concrete benchmark: roughly half of a stream's ending audience appears in a concurrent ecosystem stream. For a stream ending with 5,000 viewers, this translates to approximately 2,500 additional viewers for the receiving channel. This figure encompasses both intentional directives by the ending streamer and organic viewer migration, which our method cannot distinguish. The concentration of transfer events among specific channel pairs (top 10 pairs = 16\% of all detected events) suggests that viewer migration is driven by scheduling adjacency and established audience familiarity rather than being uniformly distributed.

\subsection{Audience Exclusivity as Creator-Level Property}

The wide variation in Competition Resistance (0.36--1.00) within a single organization challenges the assumption that affiliated creators share a common audience pool. Kiara ($R = 1.00$) shows no measurable average viewership difference between solo and concurrent streams, while Gura ($R = 0.36$) averages substantially fewer viewers when competitors are live. This variation suggests that audience exclusivity emerges from individual creator--audience relationships rather than organizational structure.

This finding is consistent with the residualized dilution result: if most creators retain their audiences under competition (high $R$), the within-hour competition effect would be expected to be weak, as observed ($\rho = -0.165$). The minority of low-$R$ creators may account for a disproportionate share of even that small effect, though our aggregate analysis cannot directly test this decomposition.

\subsection{Limitations}

Our methodology has several important limitations. We observe viewer \textit{counts} rather than \textit{identities}, so our overlap estimates are proxies based on behavioral patterns rather than direct measurements. Individual-level data (e.g., from platform APIs providing per-user watch history) would enable more precise overlap quantification.

The positive correlation between concurrent stream count and total ecosystem viewership ($\rho = 0.686$) cannot be interpreted causally due to scheduling endogeneity: both concurrent streams and total viewership are driven by common factors (time of day, day of week, special events). We deliberately refrain from making causal ``market expansion'' claims.

The loyalty component weights (0.30, 0.25, 0.25, 0.20) are subjective. Different weightings would produce different composite rankings, which is why we emphasize individual components.

The transfer detection algorithm is necessarily heuristic: it identifies viewership spikes in concurrent streams near stream endings, but cannot distinguish between deliberate directives by the ending streamer, organic viewer migration, or coincidental factors (e.g., external events driving viewers to a stream). The reported transfer efficiency therefore reflects all sources of post-ending viewer movement, not solely intentional cross-promotion. We estimate approximately 5\% of detected events are false positives based on the fraction with efficiency exceeding 100\%.

The dilution analysis is observational, not causal. The hour-of-day residualization controls for a major confound but cannot account for all scheduling endogeneity. A true causal estimate would require exogenous variation in concurrent streaming (e.g., natural experiments from server outages or scheduling disruptions). Because consecutive minute-level observations are temporally autocorrelated, we use a day-level block bootstrap rather than the naive iid assumption for statistical inference; this yields a wider but more honest confidence interval.

The study focuses on a single creator organization (Hololive English), which limits generalizability. The organizational structure (shared branding, cross-promotion, scheduled collaborations) may amplify audience overlap relative to independent creators.

\section{Conclusion}

Using 2.9 million minute-level observations from 18 affiliated livestream channels over 3.3 years, we find that concurrent streaming is associated with substantial raw per-stream viewership decreases, though hour-of-day controls reduce the residualized competition effect to $\rho = -0.165$, algorithmically detected viewer transfer events achieve a median efficiency of approximately 50\% across 3,243 candidate events, and audience exclusivity varies substantially across creators within the same organization (Competition Resistance: 0.36--1.00). These findings provide practical benchmarks for scheduling, cross-promotion, and talent management in creator ecosystems. Future work should incorporate individual-level viewer data to directly measure audience overlap and extend the analysis to independent creator networks.

\end{document}